\documentclass[preprintnumbers,amsmath,amssymb,floatfix,14pt,onecolumn,prd,superscriptaddress,nofootinbib,showpacs,showkeys]{revtex4}
\usepackage{graphicx}
\usepackage{epsfig}
\usepackage{bm}
\usepackage{amsfonts}

\def\ben{\begin{equation}}
\def\een{\end{equation}}

 \def\bd{\begin{document}} \def\ed{\end{document}}
\def\ds{\documentstyle} \let\fr=\frac \let\bl=\bigl \let\br=\bigr
\let\Br=\Bigr \let\Bl=\Bigl
\let\bm=\bibitem
\let\na=\nabla
\let\pa=\partial \let\ov=\overline
\newcommand{\be}{\begin{equation}}
\newcommand{\ee}{\end{equation}}
\def\ba{\begin{array}}
\def\ea{\end{array}}
\def\ft#1#2{{\textstyle{\frac{\scriptstyle #1}{\scriptstyle #2} } }}
\def\fft#1#2{{\frac{#1}{#2}}}
\def\del{\partial}
\def\vp{\varphi}
\def\sst#1{{\scriptscriptstyle #1}}
\def\oneone{\rlap 1\mkern4mu{\rm l}}
\def\td{\tilde}
\def\wtd{\widetilde}
\def\ie{{\it i.e.\ }}
\def\dalemb#1#2{{\vbox{\hrule height .#2pt
        \hbox{\vrule width.#2pt height#1pt \kern#1pt
                \vrule width.#2pt}
        \hrule height.#2pt}}}
\def\square{\mathord{\dalemb{6.8}{7}\hbox{\hskip1pt}}}
\newcommand{\ho}[1]{$\, ^{#1}$}
\newcommand{\hoch}[1]{$\, ^{#1}$}
\newcommand{\bea}{\setlength\arraycolsep{2pt} \begin{eqnarray}}
\newcommand{\eea}{\end{eqnarray}}
\newcommand{\ra}{\rightarrow}
\newcommand{\lra}{\longrightarrow}
\newcommand{\Lra}{\Leftrightarrow}
\newcommand{\bp}{\tilde \beta^\prime}
\newcommand{\tr}{{\rm tr} }
\newcommand{\Tr}{{\rm Tr} }
\def\0{{\sst{(0)}}}
\def\1{{\sst{(1)}}}
\def\2{{\sst{(2)}}}
\def\3{{\sst{(3)}}}
\def\4{{\sst{(4)}}}
\def\5{{\sst{(5)}}}
\def\6{{\sst{(6)}}}
\def\7{{\sst{(7)}}}
\def\8{{\sst{(8)}}}
\def\m{{\sst{(m)}}}
\def\n{{\sst{(n)}}}
\def\cA{{{\cal A}}}
\def\cB{{{\cal B}}}
\def\cF{{{\cal F}}}
\def\cG{{{\cal G}}}
\def\cH{{{\cal H}}}
\def\tV{\widetilde V}
\def\tW{\widetilde W}
\def\tH{\widetilde H}
\def\tE{\widetilde E}
\def\tF{\widetilde F}
\def\tA{\widetilde A}
\def\im{{{\rm i}}}
\def\tY{{{\wtd Y}}}
\def\ep{{\epsilon}}
\def\vep{{\varepsilon}}
\def\bD{{{\bar D}}}
\def\R{{{\mathbb R}}}
\def\C{{{\mathbb C}}}
\def\H{{{\mathbb H}}}
\def\CP{{{\mathbb C}{\mathbb P}}}
\def\RP{{{\mathbb R}{\mathbb P}}}
\def\Z{{{\mathbb Z}}}
\def\bA{{{\mathbb A}}}
\def\bB{{{\mathbb B}}}
\def\bC{{{\mathbb C}}}
\def\bD{{{\mathbb D}}}
\def\bE{{{\mathbb E}}}
\def\bZ{{{\mathbb Z}}}
\def\Re{{{\frak{Re}}}}
\def\Im{{{\frak{Im}}}}
\def\cosec{{\,\hbox{cosec}\,}}
\def\Gm{{\Gamma_{\!\! -}}}
\def\Gp{{\Gamma_{\!\! +}}}
\def\stan{{standard }}
\def\nonstan{{supernumerary }}
\def\p{{\partial}}
\def\kdel#1{{\fft{\del}{\del#1}}}

\def\bog{{Bogomolny }}
\def\om{{\omega}}

\newcommand{\nnr}{\nonumber \\}
\newcommand{\pd}{\partial}
\newcommand{\ud}{\textrm{d}}
\newcommand{\dTH}{T^{\prime \, 0}_\textrm{H}}
\newcommand{\dOi}{\Omega^{\prime \, 0}_i}
\newcommand{\bx}{{\bf x}}
\begin{document}
\title{Entropic corrections to Newton's law }
\author{\textbf{M. R. Setare}}
\email{rezakord@ipm.ir} \affiliation{Department of Science, Payame
Noor University, Bijar, Iran}

\author{\textbf{ D. Momeni}}
\email{d.momeni@yahoo.com; davidmathphys@yahoo.co.uk}
 \affiliation{Eurasian International Center
for Theoretical Physics, Eurasian National University, Astana
010008, Kazakhstan}
\author{\textbf{ R. Myrzakulov}}
\email{rmyrzakulov@gmail.com;
rmyrzakulov@csufresno.edu}
\affiliation{Eurasian International Center
for Theoretical Physics, Eurasian National University, Astana
010008, Kazakhstan}
\begin{abstract}
In this short letter we calculate separately the generalized uncertainty principle (GUP)   and self
gravitational corrections to the Newton's gravitational formula. We
show that for a complete description of the GUP and self-gravity
effects, both temperature and the entropy must be modified. 
 \end{abstract}
\keywords{Generalized uncertainty principle; Black Hole Physics; Thermodynamics of Black Holes}
\pacs{04.20.Cv 04.50.-h 04.70.Dy}
\newpage
 \maketitle
\section{Introduction}

The Planck scale corrections to the new scenario for gravity,
proposed by the Verlinde\cite{Verlinde} must contains two group of
the modifications. First the modification in the amount of the
entropy which disturbed when the test particle approach the
holographic screen. This excess correction arisen since the full
quantum description of the test particle in the Planck scale
described by the generalized uncertainty principle (GUP) and not the
usual uncertainty Principle. This work was done by Gosh \cite{Ghosh}
and Vancea \cite{Vancea} and other authors \cite{Modesto}. Attend
that in this treatment the usual Bekenstein-Hawking famous formula
for the entropy \cite{BH} preserved. But as it was shown in \cite{cav,
Setare1, Setare2} in the limit of the GUP both the temperature and
 entropy need serious modifications. This kind of the modifications
 in literature called the quantum corrections. Some authors inserted
 the new functions for the entropy versus area but no change has
 been  done on the form of the temperature. The kind of models which
 used from the various kinds of the entropy functions belongs to
 this category. One of this kind of modifications in the entropy
 comes directly from the Loop Quantum Gravity
(LQG)\cite{LQG} which this form of the modified entropy functions
orthodox carries some log terms and also an inverse term of the
area. In this approach the only change which one must perform is the
replacing the simple term $\frac{\delta S }{\delta A}$ the new term
$\frac{\partial S}{\partial A}\frac{\delta S}{\delta A}$. This work
was done by some authors \cite{Sheykhi},\cite{Modesto}. Especially
the last authors show that these modification of the simple Verlinde
idea adjust completely with MOND \cite{MOND}. Albeit there are other
alternatives for solving the dark matter problem for example the
Cosmological special relativity belong to the Carmeli\cite{Carmeli},
beyond the contemporary cosmology. Anyway in both this approaches
one important notion missed: the modification of the
temperature(Hawking or Unruh)in Planck scale. As was shown by
\cite{cav, Setare1, Setare2}, applying the GUP modifies the standard
expression for
the Hawking temperature.\\
Another corrections to the entropy function whatever the usual black
hole(BH) entropy or the entropy is the back reaction effect or the
self-gravitational corrections \cite{self}. We know that near the
horizon of any BH the spectrum of particles created which likes the
usual Planck black body spectrum. This kind of radiation changes the
background slightly in large scale but very significant near the
horizon. As was shown \cite{Das, Setare3} small statistical
fluctuations around equilibrium,changes the entropy by some
logarithmic terms. In this letter we present the full GUP inspired
corrections to the Newtonian gravity and also the self gravitational
corrections or thermal radiation corrections separately. 

\section{GUP corrections to the Hawking temperature and black hole
entropy}
 As was shown in \cite{cav, Setare1,Setare2} if one applies GUP
when the phase space which describes the motion of the particle be
of order the Planck length and the mass of it in order of the Planck
mass, the corrected formula for Hawking temperature for  a
d-dimensional Schwarzschild black hole is in the form:
\be
T'=T(1+\alpha^2\gamma T^2+O(\alpha^4))\label{gupt}
\ee
attending that one we write in (\ref{gupt}) is slightly different from which
is in  the original work. Whatever is done is the eliminating the
mass in Planck units $m=\frac{M}{M_{p}}$ between the temperature and
the parameter $\lambda_{p}$ in the original work. The not obvious
form of the constant for a d-dimensional Schwarzschild metric was
given by $\gamma=(\frac{2\pi}{(d-3)M_{p}c^2})^2$. Here T is the usual
Hawking temperature in the absence of the GUP. The parameter $\alpha$ is a
dimensionless constant of order one that controls the  potency  of
the GUP's confine. Too we assume that this coupling term is very
small and we can carry out the Taylor expansion up to 2'nd order
w.r.t  $\alpha$. The generalized uncertainty principle corrected black hole
entropy is
\be
\acute{S}=S+\alpha^2\gamma'S^{\frac{d-4}{d-2}}+O(\alpha^4)
\ee
Where for a d-dimensional schwarzschild BH  the parameter$\gamma'=constant$. For
$d=4$ the GUP correction term up to order $\alpha^2$ is a constant
term. Thus we can ignore from it. The same conclusion is obtained by
this fact that in a schwarzschild spacetime
always  $T\propto\frac{1}{\sqrt{S}}$ thus from (\ref{gupt})  we can show that up
to order 2, the correction is a constant term which can be
completely layaway in derivation of the entropic force.

\section{The generalized entropic corrections  to Newton's law}
In this section first we discuss only the GUP corrections and then
we treat the Self-Gravitational Corrections separately.

\subsection{GUP  corrections}
We accept (\ref{gupt}) as the correct expression for the Hawking temperature
near the validity of GUP.  Assume that the total change of the
screen's entropy is
\begin{eqnarray}
\delta S=\frac{\partial S}{\partial A}\delta A
\end{eqnarray}
and as we know that the infinitesimal displacement  of the test
particle with mass  $m$   must be in order of the Compton wavelength
\begin{eqnarray}
\delta x=\eta\lambda_{c}=\frac{\eta\hbar}{mc}
\end{eqnarray}
we know that the quantized holographic screen or as it was been in the
Verlinde scenario, the area of the horizon of a black hole(spherical
)must be wrought from an amount of the information bytes which is
related directly to the degree of the freedom of the horizon
\begin{eqnarray}
A=QN
\end{eqnarray}
 Here N is the number of bytes and Q is of order the planck length. The
common formula for the Newton's gravity ,assuming the modified GUP
essence formula (\ref{gupt}) for temperature, is
\begin{eqnarray}
F=T'\frac{\delta S'}{\delta x}=T'\frac{\partial S'}{\partial
A}\frac{\delta A}{\delta x}
\end{eqnarray}
Remembering (4,5)
\begin{eqnarray}
F=T(1+\alpha^2\gamma T^2+O(\alpha^4))\frac{\partial S'}{\partial
A}\frac{Q\delta N}{\eta \lambda_{c}}
\end{eqnarray}
In (7) T may be read  as the Hawking temperature for horizon or the
Unruh temperature\cite{Unruh} for an accelerated test particle (a
test particle near the horizon senses himself in a thermal bath as a
Rindler observer). Albeit in general, these different temperatures
measured by different observers are not equal. According to the
equipartition law of energy which is valid even in the GUP regime,
and the Einstein equivalency between mass and energy for T we have
\begin{eqnarray}
T=\frac{2Mc^2Q}{4\pi k_{B} R^2}
\end{eqnarray}
substituting (1,4,5,8) in (7) and by assuming that $\delta N=1$,we
have:
\begin{eqnarray}
F=F_{0}(1+\frac{\alpha^2B}{R^4}+O(\alpha^4))\label{fgup}
\end{eqnarray}
Where the $F_{0}=\frac{Mm}{R^2}\frac{Q^2c^3}{8\pi k_{B} \hbar\eta}$,
and $B=\gamma(\frac{2Mc^2Q}{4\pi k_{B}})^2$. Remembering
$l_{p}^2=\frac{G\hbar}{c^3}$ and by comparing with Newtonian force
$F_{0}$, we observe that
\begin{eqnarray}
Q=\sqrt{8\pi k_{B}\eta} l_{p}^2
\end{eqnarray}
which fixed it's value. In (\ref{fgup}) the first term is the usual newtonian
term,the second term is completely  new and directly back to the
nature of the GUP modifications of the temperature(horizon)and
entropy.

\section{Self-Gravitational Corrections to entropic force}
 First we overview the form of the the modified
temperature due to the self-gravitational effect following the idea
of Keski-Vakkuri, Kraus and Wilczek (KKW) \cite{self}. The KKW
analysis means the total energy of the spacetime under study is kept
fixed while the black hole mass is allowed to vary. We therefore
expect a black hole of initial mass M to have a final mass of $M
+\omega$ where $\omega$ is the energy of the emitted particle
\cite{SV}. Since for a Schwarzschild black hole
\begin{eqnarray}\nonumber
\frac{1}{M}=\zeta S^{-1/2}
\end{eqnarray}

self- gravitational corrections to second order in entropy for a
schwarzschild metric to be considered\cite{Setare4}
\begin{eqnarray}
S'=S(1+\frac{2\omega\zeta}{\sqrt{S}}+\frac{\omega^2\zeta^2}{S}+O(\omega^3))
\end{eqnarray}
where $\omega$ is the energy of the tunnelling particle and $\zeta$
some proportionality constant.Since we know that in a Schwarzschild
spacetime always

\begin{eqnarray}\nonumber
T=\frac{\kappa}{\sqrt{S}}
\end{eqnarray}
,thus the corrected temperature from the self-gravitational effect
is:
\begin{eqnarray}
T'=T(1-\frac{\omega\zeta}{\kappa}
T+\frac{\omega^2\zeta^2}{\kappa^2}T^2+O(\omega^3))
\end{eqnarray}
Now we are ready to calculate the self gravitational corrections for
Newtonian gravity in verlinde approach.
 Substituting (11), (12)
in (6) and carried the simple algebra we obtain:
\begin{eqnarray}
F=F_{0}(1+O(\omega^3)),
\end{eqnarray}
where $F_{0}=\frac{mM}{R^2}\frac{c^6 Q^2}{8\pi k_B G
\hbar^{2}\eta}$. Again by comparing with Newtonian force $F_{0}$, we
observe that
\begin{eqnarray}
Q=\sqrt{8\pi k_{B}\eta} l_{p}^2
\end{eqnarray}
 thus in second order of the $\omega$ there is no correction to
the Newtonian force.

\section{Summary}
There has been much recent interest in calculating the quantum
corrections to $S_{BH}$ (the Bekenestein-Hawking entropy) . The
leading-order correction is proportional to $\log (S_{BH})$. There
are, two distinct and separable sources for this logarithmic
correction. Firstly, there should be a correction to the number of
microstates that is a quantum correction to the microcanonical
entropy, secondly, as any black hole will typically exchange heat or
matter with its surrounding, there should also be a correction due
to thermal fluctuations in the horizon area. In this short letter we
show that for attaining  to a complete generalization of the
entropic corrections to the Newton's gravity, we must consider
temperature and entropy both modified in any of two different
regimes GUP and the self-gravitational corrections. These
corrections are given by Eqs.(9), (13) respectively.


\begin{thebibliography}{35}
\bibitem{Verlinde}
E. P. Verlinde,JHEP 1104:029,2011.
\bibitem{Ghosh}
S. Ghosh, 1003.0285 [hep-th].
\bibitem{Vancea}
I. V. Vancea, M. A. Santos,Mod.Phys.Lett. A27 (2012) 1250012.
\bibitem{Modesto}
L. Modesto,  A. Randono,1003.1998 [hep-th].
\bibitem{BH}
 J. D. Bekenstein, Lett. Nuovo. Cim. 4, 737 (1972); Phys. Rev. D7, 2333 (1973); Phys.
Rev. D9, 3292 (1974); S. W. Hawking, Comm. Math. Phys. 25, 152
(1972); J. M. Bardeen, B. Carter and S. W. Hawking, Comm. Math.
Phys. 31, 161 (1973).
\bibitem{cav}
M. Cavaglia and S. Das, hep-th/0404050; S. Das,hep-th/0403202.
\bibitem{Setare1}
M. R. Setare, Phys.Rev. D70 (2004) 087501
\bibitem{Setare2}
M. R. Setare, Int. J.Mod. Phys. A21 (2006) 1325-1332 .
\bibitem{LQG}
K. Krasnov, C. Rovelli, Class.
Quant. Grav. 26, 245009 (2009) ; C. Rovelli, Phys. Rev.
Lett.77:3288- 3291 (1996); O. Dreyer, F.
Markopoulou, L. Smolin,Nucl.Phys. B744 (2006).
\bibitem{Sheykhi}
 A. Sheykhi,Phys.Rev.D81:104011,2010.
 \bibitem{MOND}
M. Milgrom, Astrophys. J. 270, 371 (1983); 270, 384 (1983); 270, 365
(1983).
\bibitem{Carmeli}
S. Behar, M. Carmeli ,Int.J.Theor.Phys. 39 (2000) 1397-1404 ; M. Carmeli, Int.J.Theor.Phys. 38 (1999) 1993
; M. Carmeli,Int.J.Theor.Phys. 37 (1998) 2621-2625.
\bibitem{self} S.W. Hawking, Commun. Math. Phys. 43, 199 (1975) ; E. K. Vakkuri and P.
Kraus, Phys. Rev. D54, 7407, (1996) ; M.K. Parikh and F. Wilczek,
Phys. Rev. Lett. 85, 5042, (2000); S. Hemming and E. Keski-Vakkuri,
Phys. Rev. D64, 044006, (2001); M. R. Setare, E. C.Vagenas, Phys. Lett. B584, 127, (2004).
\bibitem{Das}
S. Das,P. Majumdar,R. K. Bhaduri,Class.Quant.Grav. 19 (2002) 2355-2368.
\bibitem{Setare3}
M.R. Setare,  Phys. Lett. B573 (2003) 173-180.
\bibitem{Unruh}
Unruh, W. G, Phys. Rev. D14, 1976,
870.
\bibitem{SV}
M.R. Setare , E. C. Vagenas, Int. J. Mod. Phys. A20 (2005)
7219-7232.
\bibitem{Setare4}
M.R. Setare, Int. J. Mod. Phys. A23:2047, (2008).
\bibitem{cai}
R.G. Cai, L. M. Cao and N. Ohta, Phys. Rev. D 81,
061501(R) (2010).



\end{thebibliography}
\end{document}